# Comment on "The GSI method for studying neutrino mass differences - For Pedestrians"

Murray Peshkin*

*Physics Division, Argonne National Laboratory, Argonne, IL 60439*

Abstract

It has been suggested that the reported temporal oscillations in the weak decay of H-like ions circulating in the GSI storage ring may be accounted for by interference between two different momentum components of the wave function of the parent ions. In that model, the interference is said to come about through coupling of those two momentum components to the two mass components in the wave function of the electron neutrino in the decayed state. I show here that quantum mechanics allows no such interference to contribute to the reported oscillation of the decay rate. The central issue is that the storage ring is a million times too small to allow the parent ion's wave function to produce the needed interference effect. This note supersedes earlier arXiv postings that were less general and less rigorous.

## 1. Introduction

In a recent experiment, hydrogen-like ions were injected into the GSI storage ring and the rate of their decay by electron capture with the emission of a neutrino was measured [1]. In the case of $^{142}$Pr ions, the measured decay rate $R(t)$ was reported to be described by

$$R(t) = R(0)e^{-\gamma t}\left[1 + a\cos(\omega t + \phi)\right] \qquad (1)$$

with $\omega = 0.9$ sec$^{-1}$. For present purposes it is sufficient to have only a rough value of $\omega$ plus the information that $\omega$ is significantly greater than $\gamma$ so that several oscillations can be seen. $R(t)$ was measured by timing successive passages of the ion past a fixed point in the storage ring. A small change in the time to circuit the ring signaled a change in the mass of the ion from that of the parent to that of the daughter. The only additional detail of the experiment that is important for what follows is that the uncertainty in the time of decay was tens of μsec, many times the orbital period of the ion, with the result that where in the ring the decay occurred was completely unknown. In other words, the quantity measured was the total decay rate $R(t)$, not the rate of decays in a limited part of the storage ring.

H.J. Lipkin [2,3] has suggested that two different momentum components in the wave function of the parent ion can feed the same final state by coupling coherently to the two different mass states of the electron neutrino. In that model, the relative phase of the those two contributions to the amplitude for decay may cycle with frequency ω and cause or at least contribute to the observed oscillation. It will be shown here that no such cycling of the relative phase is possible under the conditions of the GSI experiment.

This report supersedes two earlier arXiv postings [4,5] which reached the same conclusion that the model of Refs.[2,3] cannot account for the decay oscillations, but not generally or rigorously. Here I use only general principles of quantum mechanics plus one observation about the scales of the relevant physical quantities in the GSI experiment. The key point of the proof is that the support of the parent ion's spatial wave function is constrained by the size of the storage ring. Then its Fourier transform, the momentum-space wave function, cannot vary appreciably with momentum shifts much less than the inverse of the size of the ring, around $10^{-4}$ cm$^{-1}$. But the hypothesized interference mechanism of Refs. [2,3] would need significant ripples in the momentum-space wave function on a scale of $10^{-10}$ cm$^{-1}$ to produce the temporal oscillations reported in the experiment. Therefore the model cannot account for the experiment.

All that is proved in Section 2 below. That proof assumes an idealized model of the experiment, but it is shown in Section 3 that the idealized model can safely be applied to the real experiment.

It is also noted in Section 3 that nothing said here implies a contradiction between quantum mechanics and the reported experimental oscillations. Only the model of Refs. [2,3], in which the phenomenon arises from interference between certain momentum components of the parent ion's wave function, is challenged here.

## 2. Proof

Let the Hamiltonian for the system be

$$H = H_0 + H', \tag{2}$$

where $H'$ is the weak interaction and $H_0$ contains the energies of the particles and their interactions with electromagnetic fields in the storage ring. Those external fields may be space dependent and time dependent. The total wave function $\Psi(t)$ is given by

$$\Psi(t) = \psi(t) + \varphi(t), \tag{3}$$

where $\psi$ is the part of the wave function in which the parent ion is present and $\varphi$ is the part in which the daughter ion and a neutrino are present.

$$i\dot{\Psi}(t) = H\Psi(t) \tag{4a}$$
$$i\dot{\psi}(t) = H_0\psi(t) + H'\varphi(t) \tag{4b}$$
$$i\dot{\varphi}(t) = H_0\varphi(t) + H'\psi(t) \tag{4c}$$

The rate of decay by K-electron capture with emission of an electron neutrino is given by

$$R(t) = \frac{d}{dt}\langle\varphi(t)|\varphi(t)\rangle = 2\,\text{Im}\{\langle\varphi(t)|H'|\psi(t)\rangle\} \tag{5}$$

The wave functions can be written as sum/integrals over orthogonal basis sets as

$$\psi(t) = \int dM\,d\mathbf{p}\,|A,\mathbf{p},M\rangle\langle A,\mathbf{p},M|\psi(t)\rangle \tag{6a}$$
$$\varphi(t) = \sum_m \int d\mathbf{p}\,d\mathbf{q}\,|D,\mathbf{p},\mathbf{q},m\rangle\langle D,\mathbf{p},\mathbf{q},m|\varphi(t)\rangle \tag{6b}$$

Here, $|D,\mathbf{p},\mathbf{q},m\rangle$ are the basis states for the decayed system with total momentum $\mathbf{p}$, relative momentum of the neutrino and the daughter nucleus $\mathbf{q}$, and neutrino mass $m$; $|A,\mathbf{p},M\rangle$ are the basis states for the parent nucleus with momentum $\mathbf{p}$ and mass $M$. For a stable nucleus, $M$ would be a number, but for a decaying state $M$ is a continuous eigenvalue of the mass operator. The dependence of the wave function $\langle A,\mathbf{p},M|\psi(t)\rangle$ upon $M$ is presumably something like a Breit-Wigner shape or perhaps in this case a sum of two or more such shapes, but that information will not be used here. The basis states are generally not eigenstates of $H_0$.

Inserting Eqs.(6) into Eq.(5) gives

$$R(t) = 2\,\text{Im}\left\{\sum_m \int d\mathbf{p}\,d\mathbf{q}\,d\mathbf{p}'\,dM\,\varphi(\mathbf{p},\mathbf{q},m,t)^*\langle D,\mathbf{p},\mathbf{q},m|H'|A,\mathbf{p}',M\rangle\psi(\mathbf{p}',M,t)\right\} \tag{7}$$

where $\psi(\mathbf{p},M,t)$ and $\varphi(\mathbf{p},\mathbf{q},m,t)$ are abbreviations for the momentum-representation wave functions $\langle A,\mathbf{p},M|\psi(t)\rangle$ and $\langle D,\mathbf{p},\mathbf{q},m|\varphi(t)\rangle$, rsp. From Eq.(7), it is apparent that final states of any momentum $\mathbf{p}$ can in principle be fed by parent states that do not all have that same momentum. An exception is the case in which there are no external fields so that momentum is conserved. In that case, $\langle D,\mathbf{p},\mathbf{q},m|H'|A,\mathbf{p}',M\rangle \propto \delta(\mathbf{p}-\mathbf{p}')$ and there is no possibility of different momentum components of the parent ion's wave function feeding the same momentum component of the final state. However that does not apply to ions moving in a storage ring.

Eq.(7) for $R(t)$ is a consequence of exact quantum mechanics, but its useful application to motion in a storage ring requires the approximation that $\psi(\mathbf{p}',M,t)=\psi(\mathbf{p},M,t)$ for all relevant values of $\mathbf{P}=\mathbf{p}'-\mathbf{p}$. In units with $\hbar=1$,

$$\psi(\mathbf{p},M,t) = \int d\mathbf{r}\,\tilde{\psi}(\mathbf{r},M,t)e^{-i\mathbf{r}\cdot\mathbf{p}} \tag{8a}$$
$$\psi(\mathbf{p}',M,t) = \int d\mathbf{r}\,\tilde{\psi}(\mathbf{r},M,t)e^{-i\mathbf{r}\cdot\mathbf{p}}e^{-i\mathbf{r}\cdot\mathbf{P}}, \tag{8b}$$

$\tilde{\psi}(\mathbf{r},M,t)$ being the configuration space wave function $\langle A,\mathbf{r},M|\psi(t)\rangle$.

For $\psi(\mathbf{p})$ and $\psi(\mathbf{p}')$ to have a relative phase that cycles with a frequency near the measured oscillation frequency $\omega$, it is necessary that the momentum shift by the amount $\mathbf{P}$ shifts the energy by an amount $\Delta E \sim \omega \sim 1\ \text{sec}^{-1}$. For relativistic particles, the $\mathbf{P}$ necessary to get those frequencies will have magnitude on the order of $P \sim \Delta E/c \sim 10^{-10}$ cm$^{-1}$. The values of **r** for which the spatial wave function $\tilde{\psi}$ is non vanishing is limited by the size of the storage ring, $\sim 10^4$ cm. Therefore, $\mathbf{r} \cdot \mathbf{P}$ in Eq.(8b) is limited to values around $10^{-6}$ and, to an accuracy of parts in a million,

$$R(t) = 2\operatorname{Im}\left\{\sum_m \int d\mathbf{p}\,d\mathbf{q}\,d\mathbf{p}'\,dM\,\varphi(\mathbf{p},\mathbf{q},m,t)\langle D,\mathbf{p},\mathbf{q},m|H'|A,\mathbf{p}',M\rangle\psi(\mathbf{p},M,t)\right\} \qquad (9)$$

Eq.(9) does not say that different values of the parent ion's momentum do not contribute to the same momentum in the final state. They do contribute. But their relative phase is always unity; it cannot vary over time with the frequency of the observed oscillations in the decay rate $R(t)$.

### 3. Conclusions

It was proved above that the reported decay oscillations in the GSI experiment cannot arise from oscillating interference between the contributions of different momenta in the wave function of the parent ion. That does not exclude the possibility of decay oscillations arising from interference between one value of $M$ in the parent ion feeding one neutrino mass state $m$ and a different $M$ feeding the other neutrino mass state. However, the suggested interference between the contributions of different momentum values has nothing to do with it.

It was also noted that a stronger statement can be made about the case where the ions move in a field-free space. There, interference between different momentum components in the wave function of the parent ion cannot cause even a constant contribution to the decay rate.

The proof given here invoked three harmless idealizations. For simplicity, the state of the parent ion was described as a pure one, represented by a wave function. In reality, that state is a statistical mixture of pure states. The proof given above applies separately to each pure state and therefore to the mixture.

In reality, the wave function, or more generally the density matrix of both the parent state and the decayed state, may be altered by interaction with the detector used for timing the circuit around the storage ring, or by electron cooling in one part of the ring. That does not matter. Eq.(5) for $R(t)$ requires only that Eq.(4c) be obeyed for an infinitesimal time around $t$. How the wave function came to have the value that it had just before that time is immaterial.

In Eqs.(4), it was assumed that there is no other decay channel than the one in which neutrino emission proceeds through electron capture. In reality, a positron may be emitted in competition with the capture of the orbiting electron. To include that channel, an additional term must be added to the equation for $\dot{\Psi}$. However, that leaves Eq.(4c), the only one used in calculating $R(t)$, unchanged.

**Acknowledgements**

This work was supported by the U.S. Department of Energy, Office of Nuclear Physics, under Contract No. DE-AC02-06CHI1357. I thank Fritz Coester for reading the manuscript and helping me clarify the presentation.

**References:**
[1] Yu. A. Litvinov *et al*, Phys. Lett. B, **664**, 162 (2008)
[2] H.J. Lipkin, arXiv:0801.1465v2 [hep-ph]
[3] H.J. Lipkin, arXiv: 0805.0435v2 [hep-ph]
[4] M. Peshkin, arXiv:0803.0935 [hep-ph]
[5] M. Peshkin, arXiv:0804.4891 [hep-ph]

*email: peshkin@anl.gov